\documentclass[epj,referee]{svjour}
\usepackage{textcomp}
\usepackage{epsfig}

\begin{document}
\begin{sloppy}

\title{Lattice structure and magnetization of
LaCoO$_{3}$ thin films}

\author{A.~D. Rata\thanks{\emph{Corresponding author:}
email d.rata@ifw-dresden.de}, A. Herklotz, L. Schultz, and K.
D\"{o}rr }
%
%
\institute{IFW Dresden, Institute for Metallic Materials,
Helmholtzstra$\ss$e 20, 01069 Dresden, Germany }
\date{Received: date / Revised version: date}

\abstract{We investigate the structure and magnetic properties of
thin films of the LaCoO$_{3}$ compound. Thin films are deposited
by pulsed laser deposition on various substrates in order to tune
the strain from compressive to tensile. Single-phase (001)
oriented LaCoO$_{3}$ layers were grown on all substrates despite
large misfits. The tetragonal distortion of the films covers a
wide range from $-2\%$ to $2.8\%$. Our LaCoO$_{3}$ films are
ferromagnetic with Curie temperature around 85~K, contrary to the
bulk. The total magnetic moment is below $1\mu_{B}$/Co$^{3+}$, a
value relatively small for an exited spin-state of the Co$^{3+}$
ions, but comparable to values reported in literature. A
correlation of strain states and magnetic moment of Co$^{3+}$ ions
in LaCoO$_{3}$ thin films is observed.
} 
\authorrunning{A.~D. Rata et al.}
\titlerunning{Lattice structure and magnetization of
LaCoO$_{3}$ thin films}
\maketitle
\thispagestyle{empty}

\section{Introduction}
The nature of the phase transition observed in LaCoO$_3$ compound
from a nonmagnetic insulator at low temperatures to a paramagnetic
semiconductor above 90~K has triggered tremendous experimental and
theoretical efforts in the last years, see Refs.
\cite{Imada,Craco08,Hozoi09} and references therein. Recently,
LaCoO$_3$ has attracted renewed interest due to the observation of
ferromagnetism in thin films
\cite{Fuchs07,Fuchs08,Herklotz09,Mehta09}. The bulk LaCoO$_3$ is
nonmagnetic at low temperatures, with Co$^{3+}$ ions having low
spin configuration (S=0). Contrary, the LaCoO$_3$ thin films grown
on various substrates exhibit a ferromagnetic ground state below
85~K \cite{Fuchs08,Herklotz09,Mehta09}. Regarding the origin of
the observed ferromagnetism, several scenarios can be found in
literature, \textit{i.e.} Jahn-Teller distortions and the rotation
of the CoO$_6$ octahedra, oxygen vacancies, increased volume and
strain-enhanced ferromagnetism
\cite{Fuchs07,Herklotz09,Mehta09,Freeland08}. On the theoretical
side, the findings are contradictory. Rondinelli \textit{et al.}
\cite{Rondinelli09} found that strain-induced changes in lattice
parameters are insufficient to cause transitions to exited spin
states of Co$^{3+}$ ions at reasonable values of strain. Gupta
\textit{et al.} \cite{Gupta09} recently indicated that the
strain-induced pseudo-tetragonal structure is responsible for the
spin-state transition in LaCoO$_3$. The authors of Ref.
\cite{Gupta09} observed that there is a very small CoO$_6$ angle
dependence on strain, contrary to earlier suggestions. Recently, a
microscopic evidence of a strain-enhanced ferromagnetic state was
reported by Park \textit{et al.}\cite{Park09}. Using magnetic
force microscopy, a ferromagnetic ground state has been confirmed
for tensile-strained films, while local magnetic clusters were
found for a relaxed film. Herklotz \textit{et al.} observed that
tetragonal distortion increases the magnetization in
tensile-strained films. The above mentioned studies agree to some
extend that the ferromagnetism in LCO is enhanced by the
substrate-induced strain and that the light hole doping alone is
not sufficient to produce ferromagnetism.

In this report, we present a detailed study of the structure and
magnetic properties of LaCoO$_{3}$ thin films grown by pulsed
laser deposition. Various substrates were used for the film growth
in order to tune the strain in the epitaxial layers from tensile
to compressive. All our LCO films are ferromagnetic with Curie
temperatures slightly varying under different strain states. The
saturated magnetic moments are much smaller than expected
(theoretical) values for an excited spin-state of the Co$^{3+}$
ions, but comparable to the previously published data. A much
stronger influence of different strain states is observed on the
variation of the total magnetic moments compared to changes
detected in the Curie temperatures. The LCO films grown under
tensile strain have the largest magnetic moments.

\section{Experiment}
LaCoO$_{3}$ thin films were grown by off-axis pulsed laser
deposition (KrF 248~nm) on various [001]- oriented
single-crystalline substrates, i.e. SrTiO$_{3}$ (STO), LaAlO$_{3}$
(LAO), (La,Sr)(Al,Ta)O3 (LSAT) and SrLaAlO$_4$ (SLAO). SiO$_2$
(thermally oxidized Si wafer) was used as well for depositing a
thick polycrystalline LCO film. All substrates were cleaned
\textit{ex-situ} with solvents and immediately introduced into the
vacuum chamber. We used stoichiometric LaCoO$_{3}$ target for
depositing our films. The deposition temperature and the oxygen
background pressure were 650$^{\circ}$C and 0.45~mbar,
respectively. After deposition, the films were annealed for
10~minutes at the deposition pressure and cooled down in oxygen
atmosphere of 800~mbar. Thickness and deposition rate were
determined from X-ray reflectivity measurements. The structure of
the LCO films was characterized \textit{ex-situ} by x-ray
diffraction (XRD). The XRD measurements were carried out with a
Philips X'Pert MRD diffractometer using Cu $K_\alpha$ radiation.
For the magnetic characterization we employed SQUID magnetometry.
Conductivity measurements were performed in the standard
four-point geometry at temperatures between 50 and 300~K. Atomic
force microscopy was employed to characterize the surface
morphology and the roughness.

\section{Results}
\subsection{Structural characterization}\label{structure}
\begin{figure}[t]
\centering{\epsfig{file=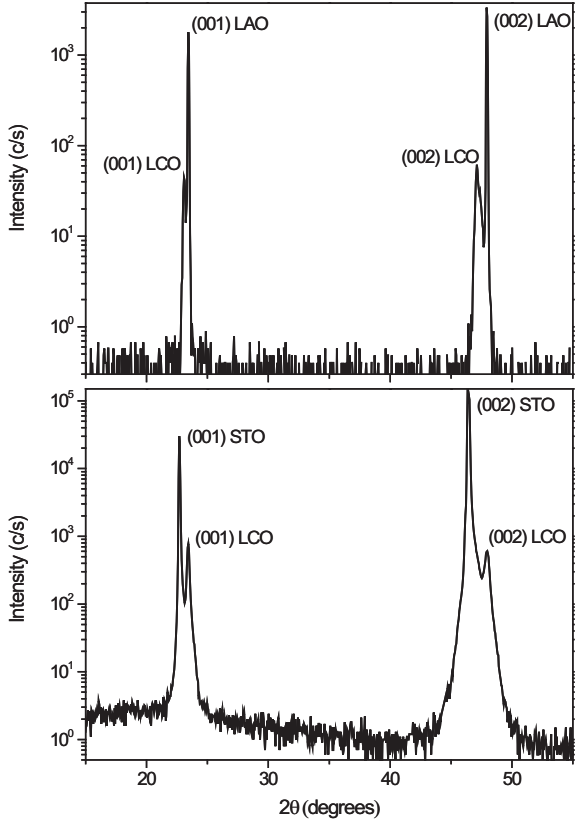,width=0.9\linewidth}}
\caption{$\theta\!-\!2\theta$ XRD scans of 100~nm-thick LCO films
deposited on LAO (top) and STO (bottom) substrates.}
\end{figure}

\begin{figure}[t]
\centering{\epsfig{file=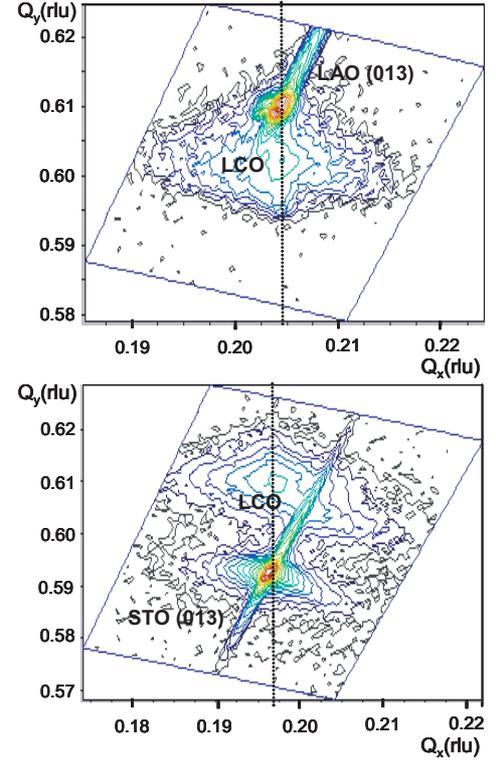,width=0.7\linewidth}}
\caption{XRD reciprocal space maps around the (013) reflection for
a LCO/LAO film (compressive strain, top) and a LCO/STO film
(tensile strain, bottom), respectively.}
\end{figure}

In the following we discuss the structural characteristics of
LaCoO$_3$ films grown on various substrates. Figure~1 shows
$\theta\!-\!2\theta$ XRD scans of 100~nm thick LCO films deposited
on (001) oriented LAO (top) and STO (bottom) single-crystalline
substrates. Both films show clear (00l) reflections of the
pseudocubic structure. No indication of impurities or
misorientation was detected. Bulk LCO, LAO and STO have the
pseudocubic lattice  parameters of 3.805, 3.79 and 3.905~$\AA$,
respectively, at room temperature.  The lattice mismatch
calculated as $(a_{s}-a_{b})/a_{s}$, where $a_{b}$  is the
pseudocubic bulk lattice parameter of LCO and $a_s$ is the lattice
parameter of the substrate is -0.4~$\%$ for LAO and +2.56~$\%$ for
STO, respectively. The LCO films grow under compressive strain on
LAO substrate, while on STO they experience a tensile strain. In
order to obtain in-plane structural information, reciprocal space
mapping (RSM) measurements around the (013) reflection were
performed. The RSM results are shown in Figure~2. LCO/LAO and
LCO/STO films are coherently grown, i.e. both films and the
substrate have the same in-plane $Q_{\mathrm{x}}$ value. The
tetragonal distortion of the films estimated as
$t\!=\!2(a-c)\!/\!(a+c)$ covers a wide range from $-2\%$~(LAO) to
$2.8\%$~(STO). The pseudocubic in-plane $a$ and out-of-plane $c$
lattice parameters and the tetragonal distortion $t$ are listed in
Table 1. The fact that 100~nm thick LCO films discussed above are
fully strained in plane is surprising, taking into account that
the critical thickness is less than 10~nm on these substrates
\cite{Mehta09}. The same behaviour was observed also for a LCO
film grown on a LSAT substrate (see Table 1). The lattice mismatch
for LCO/LSAT is $1.6\%$, much smaller than LCO/STO. Therefore, the
LCO/LSAT film is slightly tetragonally distorted compared to
LCO/STO (both grown under tensile strain).

In order to study the role played by epitaxial strain on the
occurrence of ferromagnetism in LCO films we attempt also to grow
relaxed LCO films. SLAO with an in-plane lattice parameter
$a=3.74$~$\AA$ was chosen as a substrate for the growth of relaxed
LCO films. The in-plane lattice mismatch is -1.5 $\%$, much larger
compared to LAO substrate. In figure~3 we show the results of
X-ray investigation of a 100~nm thick LCO/SLAO film. The SLAO
substrate has a tetragonal lattice. The LCO film adopt a
pseudocubic perovskite structure. In the $\theta\!-\!2\theta$ XRD
scans (see Fig.~3a) only (001) and (002) peaks of the LCO films
are visible. In-plane X-ray measurements were performed around
(107) reflection of the substrate. In the RSM map shown in Fig.~3b
the LCO (102) film reflection is also clearly visible, which
proves the epitaxial growth of the LCO film on the tetragonal SLAO
substrate. The LCO/SLAO film is almost fully relaxed, with the
$a\sim c\sim3.82$~$\AA$.

In our previous report \cite{Herklotz09}, we investigated the
magnetic properties of LCO films grown on a piezoelectric
substrate, \textit{e.g.}
Pb(Mg$_{1/3}$Nb$_{2/3}$)$_{0.72}$Ti$_{0.28}$O$_3$(001)~(PMN-PT),
which allows to explore directly the strain-dependent properties
\cite{Doerr09,Rata08,Bilani08,Dekker09}. The strain of the films
can be reversibly and uniformly controlled by the inverse
piezoelectric effect of the substrate. LCO films grown on PMN-PT
substrates are partially relaxed, \textit{i.e.} the in-plane
lattice parameter of the film is different from the substrate.
These films experience a tensile strain, the PMN-PT substrate
having a larger pseudocubic lattice parameter, i.e. 4.02~$\AA$.
The $a$ and $c$ lattice parameters and tetragonal distortion $t$
of a 100~nm thick LCO/PMN-PT film are given in Table 1. From
Table~1 one notices that all LCO films have an increased volume
compared to the bulk unit cell, regardless of the choice of the
substrate. Small angle XRR measurements show smooth and well
defined interfaces between the LCO film and the single-crystalline
substrates. Atomic force microscopy measurements also indicate a
smooth surface morphology with a roughness (rms) ranging from 0.5
to 1~nm. We note that LCO/STO films thicker than 100~nm tend to
crack, most likely caused by structural relaxation and thus the
relief of tensile strain. A last remark concerning the growth,
when SiO$_2$ was used as a substrate (not shown) the XRD spectra
indicate a polycrystalline morphology of the LCO film.

\begin{figure}[t]
\centering{\epsfig{file=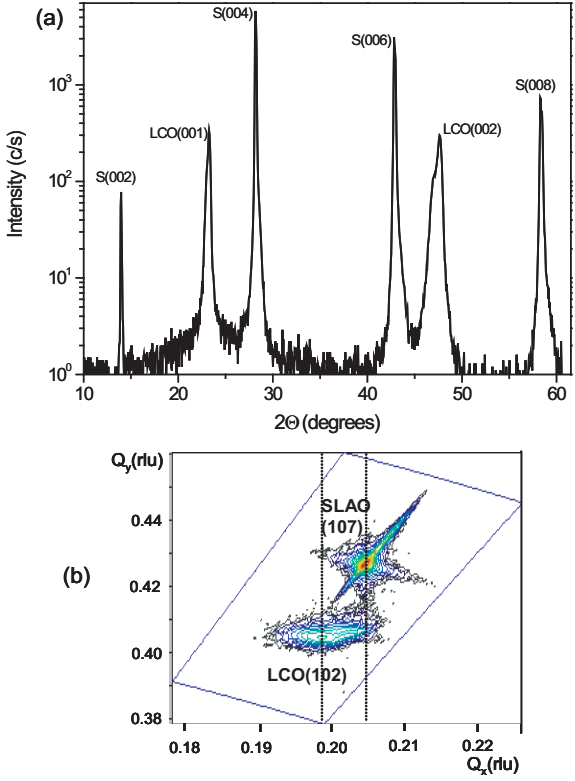,width=0.9\linewidth}}
\caption{(a) $\theta\!-\!2\theta$ XRD scans of 100~nm-thick LCO
film deposited on SLAO substrate. (b) XRD reciprocal space map of
the same film, showing the (107) reflection of the substrate and
(102) reflection of the LCO film.}
\end{figure}

\begin{table}[t]
\caption{In-plane ($a$) and out-of-plane ($c$) lattice parameters,
tetragonal distortion ($t$) and unit cell volume ($V$ = $a^2c$) of
LCO films grown on various substrates. $a_{sub}$ is the
pseudocubic substrate lattice parameter. The unit cell volume of
bulk LCO is $55.08$~$\mathrm{\AA^{3}}$.}
\begin{tabular}{lccccc}
\hline\noalign{\smallskip}
 LCO films     &$a_{sub}$ (\AA )
                           &$c$ (\AA )
                                        &$a$ (\AA )
                                                      &$t$ (\% )
                                                                   &$V$( ${\mathrm{\AA}}^{3}$ )
                                                                   \\
(100 nm)   \\

\noalign{\smallskip}\hline\noalign{\smallskip}
LCO/STO        &3.905      &3.785       &3.896       &2.8        &57.45          \\
LCO/LAO        &3.79       &3.85        &3.789       &-2         &55.27          \\
LCO/LSAT       &3.87       &3.804       &3.867       &1.6        &56.88          \\
LCO/PMN-PT     &4.02       &3.80        &3.87        &1.5        &57.06          \\
LCO/SLAO       &3.74       &3.81        &3.82        &0.2        &55.59          \\
\noalign{\smallskip}\hline
\end{tabular}
\end{table}

\subsection{Magnetic characterization}\label{magnetic}
\begin{figure}[t]
\centering{\epsfig{file=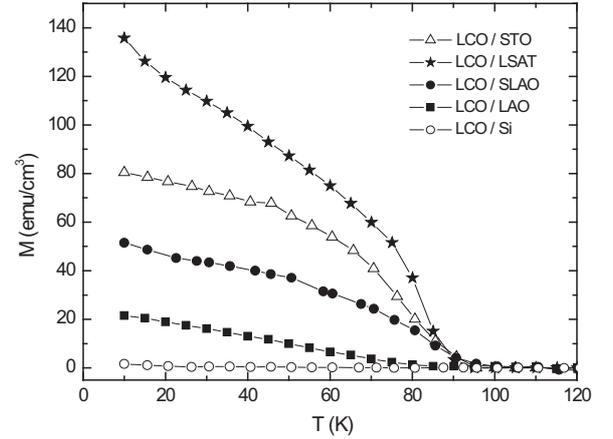,width=0.9\linewidth}} \caption{
Temperature dependence of the magnetization of various LCO films
measured after field cooling in a field of 200~mT. The Curie
temperatures are listed in Table 2.}
\end{figure}

\begin{figure}[t]
\centering{\epsfig{file=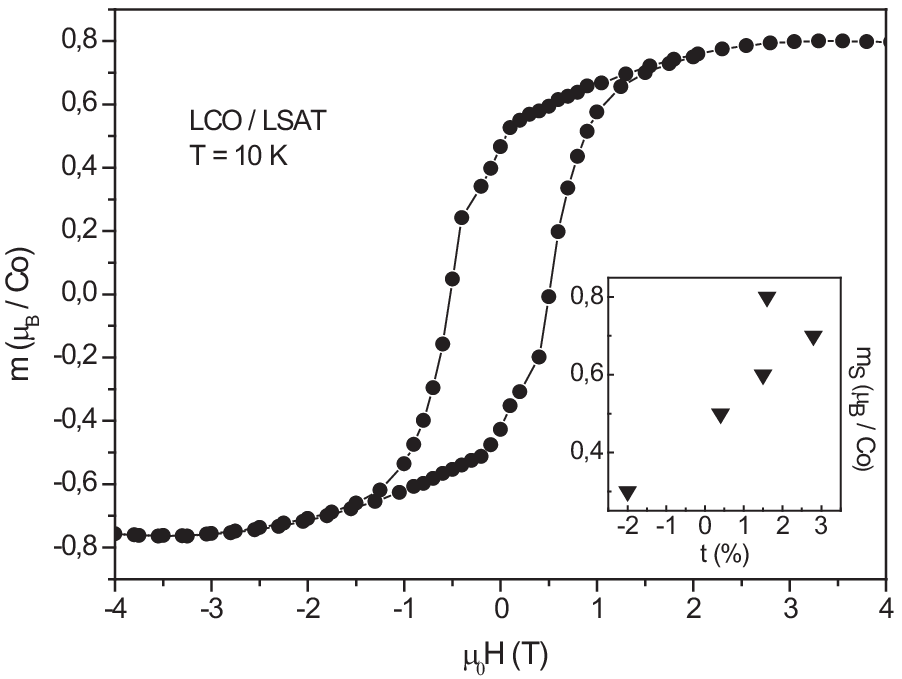,width=0.9\linewidth}}
\caption{ Field dependence of the magnetization of 100~nm thick
LCO/LSAT film measured at 10~K. Inset: variation of total magnetic
moment ($m_{S}$), measured at 10~K and $\mu_{0}H=5$~T, with
tetragonal distortion $t$ for various LCO films.}
\end{figure}

In the following the magnetic characteristics of LCO films grown
on various substrates will be discussed. In order to avoid size
effects we prepared films with the same thickness. All LCO films
grown on single-crystalline substrates display ferromagnetic
behaviour at low temperatures. On the contrary, when SiO$_2$ was
employed as a substrate, no ferromagnetic order was observed down
to 10~K. This indicates that polycrystalline LCO films resemble
the bulk behaviour, where no magnetic order is observed. In Fig.~4
we present $M$--$T$ curves of 100~nm-thick LCO films grown on
various substrates. Magnetization measurements were done after
field cooling in a field of 200~mT applied in the film plane.
Interestingly, the ferromagnetic Curie temperatures do not
strongly vary for different substrates, in spite of quite
different strain states. On the other hand, a much stronger impact
of various strain states is observed on the values of the total
magnetic moments. Field dependence of the magnetization was
measured at 10~K for all LCO films and the total magnetic moment
was estimated from the saturated magnetization. In Fig.~5 we plot
a representative $M$--$H$ loop of a 100~nm thick LCO/LSAT film.
Inset shows the variation of the Co saturation magnetic moment
($m_S$) with the tetragonal distortion $t$ for all LCO films. A
clear correlation between $m_S$ and $t$ is found, \textit{i.e.} an
increase of total magnetic moment with increasing tetragonal
distortion. Magnetic moments, Curie temperatures, and coercive
fields of all LCO films shown in Fig.~4 are compiled in Table~2.
One can notice that the values of the saturated magnetic moments
of the Co$^{3+}$ measured at 10~K in a magnetic field of 5~T are
quite small. The largest values of the saturated magnetic moment
were obtained for films grown under tensile strain. The highest
saturation magnetization observed is approximately 0.8
$\mu_{B}$/Co for a LCO/LSAT film. This is a small value but still
comparable to data reported in literature \cite{Fuchs08,Mehta09}.
If one assumes only a spin contribution to the magnetic moment, a
value of 2~$\mu_{B}$ (IS) and 4~$\mu_{B}$ (HS), respectively, is
expected. The relatively small values of the Co$^{3+}$ magnetic
moments may indicate a mixture of Co atoms in different spin
states. X-ray absorbtion experiments which could give more insight
about the Co spin state in LCO films are in progress. $T_c$ and
magnetic moments we obtained from LCO films grown under either
tensile or compressive strain are comparable to the values
reported in Refs.~\cite{Fuchs08,Mehta09} for similar thickness. It
is interesting to notice that all films have relatively large
coercive fields $H_c$ in the range from 475 to 645~mT (see
Table~2). This may indicate strong magnetocrystalline anisotropy
contributions and/or domain wall pinning, which deserve further
investigations. A clear bifurcation of zero-field cooling and
field-cooling  $M$--$T$ curves is observed (not shown) in all LCO
films which may indicate a glassy (ferromagnetic) state as found
in bulk La$_{x}$Sr$_{1-x}$CoO$_{3}$, with $x\leq0.16$ \cite{Wu06}.
Electrical transport measurements performed between 300~K and 50~K
revealed a highly insulating behaviour for all LCO films.

Despite several confirmations of ferromagnetism in LCO films from
different groups, see Refs.
\cite{Fuchs07,Fuchs08,Herklotz09,Freeland08,Mehta09}, the true
origin of the observed ferromagnetic state is not fully
understood. Although our investigations clearly indicate that
epitaxial strain plays an important role in stabilizing an exited
spin-state of the Co$^{3+}$ ions at low temperatures, some
questions remain to be answered. The tetragonal distortion induced
by epitaxial strain appears to be a key ingredient of
ferromagnetism as confirmed also by calculations \cite{Gupta09},
where the $a$ and $c$ lattice parameters have been kept fixed to
the experimental values reported in Refs.\cite{Fuchs07,Fuchs08}.
However, the occurrence of ferromagnetism in fully relaxed LCO
films grown on SLAO substrates having no tetragonal distortion
remains puzzling. Furthermore, a $T_c$ about 85~K has been also
reported in LaCoO$_3$ nanoparticles, also free of distortions
\cite{Fita08}. It is worth mentioning that a strong magnetic
signal was observed in lightly-hole-doped
La$_{1-x}$Sr$_{x}$CoO$_{3}$, x=0.002 \cite{Kataev08}, which was
attributed to the formation of spin-state polarons. The authors of
Ref. \cite{Kataev08} found that holes introduced in the low spin
state of LaCoO$_3$  by substitution of Sr$^{2+}$ for La$^{3+}$
transform the six nearest neighboring Co$^{3+}$ ions to the
intermidiate spin state forming octahedrally shaped spin-state
polarons. These spin-state polarons behave like magnetic
nanoparticles embedded in an insulating nonmagnetic matrix. The
role of light hole doping due to oxygen vacancies on the
ferromagnetism in LaCoO$_3$ films can't be totally ruled out.
However, our previous investigations \cite{Herklotz09} of LCO
films deposited on piezoelectric substrate under tensile strain
clearly show that the magnetization decreases with the reversible
release of tensile strain below $T_c$. Using this approach,
\textit{e.g.} reversible control of the strain by the inverse
piezoelectric effect of the substrate, one can overcome the effect
of other parameters, \textit{e.g.} oxygen nonstoichiometry. The
increased volume observed in all LCO films prepared on different
substrates is consistent with the occurrence of an exited state of
Co$^{3+}$ ions. The ionic radius of low-spin LS Co$^{3+}$ (0.545
$\AA$) is smaller than that of high-spin HS Co$^{3+}$ (0.61 $\AA$)
\cite{Shannon}. Finally, we note that the highly insulating
behaviour shown by all LCO films indicates that the ferromagnetic
coupling mechanism is different from Sr-doped
La$_{1-x}$Sr$_{x}$CoO$_{3}$ unless the ferromagnetism is not
long-range, but occurs in isolated clusters.

\begin{table}[t]
\caption{Magnetic moment ($m_{S}$) at 10~K, Curie temperature
($T_{C}$), and coercive field ($H_{C}$) of LCO films grown on
different substrates.}
\begin{tabular}{lccccc}
\hline\noalign{\smallskip}
LCO films     &$m_{S}$ ($\mu_{B}$/f.u.)
                           &$T_{C}$ (K)
                                        &$H_{C}$ (mT )
                                         \\
(100 nm)   \\
\noalign{\smallskip}\hline\noalign{\smallskip}
LCO/STO        &0.7      &86        &344          \\
LCO/LAO        &0.3      &75        &475          \\
LCO/LSAT       &0.8      &85        &560          \\
LCO/PMN-PT     &0.6      &87        &430          \\
LCO/SLAO       &0.5      &84        &645          \\
\noalign{\smallskip}\hline
\end{tabular}
\end{table}

\section{Summary}

To summarize, we have studied the influence of various strain
states on the magnetic properties of LaCoO$_3$ thin films.
Single-phase (001) oriented LaCoO$_{3}$ layers were grown on all
substrates despite large misfits, with tetragonal distortion
varying from $-2\%$ to $2.8\%$. All LaCoO$_{3}$ films are
ferromagnetic at low temperatures, contrary to the bulk. The Curie
temperature shows little variation with different strain states.
We observed an increase of Co saturation magnetic moment with
increasing tetragonal distortion. This result confirms that
tetragonal distortion plays an important role in the occurrence of
ferromagnetism in LaCoO$_{3}$ thin films.

\begin{acknowledgement}
We would like to thank T. Kroll, D. Fuchs and K. Nenkov for
discussions and useful suggestions. This work was supported by
Deutsche Forschungsgemeinschaft, FOR $520$.
\end{acknowledgement}

\end{sloppy}
\end{document}